\documentclass[fleqn,usenatbib]{mnras}

\usepackage{newtxtext,newtxmath}
\usepackage[T1]{fontenc}
\DeclareRobustCommand{\VAN}[3]{#2}
\let\VANthebibliography\thebibliography
\def\thebibliography{\DeclareRobustCommand{\VAN}[3]{##3}\VANthebibliography}

\usepackage{graphicx}	% Including figure files
\usepackage{amsmath}	% Advanced maths commands
\usepackage{url}

\title[Radio expansion parallax in CRL 618]{The distance to CRL~618 through its radio expansion parallax}

\author[L. Cerrigone et al.]{Luciano Cerrigone,$^{1,2}$\thanks{E-mail: luciano.cerrigone@alma.cl}
Grazia Umana,$^{3}$
Corrado Trigilio,$^{3}$
Karl M. Menten,$^{4}$
Cristobal Bordiu,$^3$ \newauthor
Adriano Ingallinera,$^{3}$
Paolo Leto,$^{3}$
Carla S. Buemi,$^{3}$
Filomena Bufano,$^{3}$
Francesco Cavallaro,$^{3}$
Sara Loru $^{3}$ \newauthor
and Simone Riggi,$^3$
\\
$^{1}$National Radio Astronomy Observatory, 520 Edgemont Road, Charlottesville, VA 22903, USA\\
$^2$Joint ALMA Observatory, Alonso de C{\'o}rdova 3107, Vitacura, Santiago, Chile\\
$^{3}$INAF-Osservatorio Astrofisico di Catania, via S. Sofia 78, 95123, Catania, Italy\\
$^{4}$Max-Planck-Institut f\"ur Radioastronomie, Auf dem H\"ugel 69. Bonn, Germany\\
}

\date{Accepted XXX. Received YYY; in original form ZZZ}

\pubyear{2023}

\begin{document}

\label{firstpage}
\pagerange{\pageref{firstpage}--\pageref{lastpage}}
\maketitle

\begin{abstract}
CRL~618 is a post-AGB star that has started to ionize its ejecta. Its central HII region has been observed over the last 40 years and has steadily increased in flux density at radio wavelengths. In this paper, we present data that we obtained with the Very Large Array in its highest frequency band (43~GHz) in 2011 and compare these with archival data in the same frequency band from 1998. By applying the so-called expansion-parallax method, we are able to estimate an expansion rate of $4.0\pm0.4$~mas~yr$^{-1}$ along the major axis of the nebula and derive a distance of $1.1\pm0.2$~kpc. Within errors, this distance estimation is in good agreement with the value of $\sim$900~pc derived from the expansion  of the optical lobes.
\end{abstract}


%%%%%%%%%%%%%%%%%%%%%%%%%%%%%%%%%%%%%%%%%%%%%%%%%%

%%%%%%%%%%%%%%%%% BODY OF PAPER %%%%%%%%%%%%%%%%%%

\section{Introduction}
The fate of a star with a main sequence mass in the range from $0.8$ to 8~M$_\odot$ is well established: it goes through the Asymptotic Giant Branch (AGB) phase, then into the Planetary Nebula (PN) phase and eventually it ends its evolution as a white dwarf \citep{kwok78}.
However, the details of each evolutionary stage, such as the formation and the early evolution of PNe are not clear. For example, it is still uncertain which mechanism/s operate to produce the often complex morphology observed in a fraction of PNe and post-AGB stars, while the circumstellar envelopes (CSEs) around their progenitors (AGB stars) show regular and symmetric structures. 

It is commonly accepted that the circumstellar shells of PNe are the product of the interaction between the residual, slowly expanding ($\sim$20~km~s$^{-1}$) AGB envelope and a subsequent, rapid (100--1000~km~s$^{-1}$) shaping agent. High-speed collimated outflows, or jets, that operate during the early post-AGB evolutionary phase, or even towards the end of the AGB, have been proposed as the primary agent in  shaping CSEs, the origin of such winds being linked to binarity \citep{soker1998, demarco}.

To investigate this matter, many authors have tried to identify very young PNe or proto-Planetary Nebulae (PPNe), but this has turned out to be quite difficult, due to the short duration ($\approx$1000 yr) of the transition from the AGB to the PN stage.
One of the few objects that can be used for such studies is CRL~618, a PPN that started its post-AGB stage about 200 years ago and is rapidly evolving towards the PN stage  {\citep{westbrook}}. The rapid evolution of this object is explained by the nature of its central object, which has been found to be an active symbiotic, whose companion star displays a WC8-type spectrum \citep{balick14}. This source provides us
 with a unique opportunity to study the physical processes taking place immediately before the birth of a planetary nebula. Multi-wavelength, multi-epoch observations \citep{sanchez02, sanchez04_1, sanchez04_2, pardo, soria} have provided us with a complex picture of the source, consisting of: 
i) a large ($\sim$20$''$) molecular envelope made of the remnant AGB wind expanding at low velocity ($\sim$17~km~s$^{-1}$); 
ii) multiple optical lobes, where shocked gas expands at high velocity  ($\sim$200 km/s);
iii) a fast (up to $\sim$340~km~s$^{-1}$) bipolar molecular outflow along the polar axis;
iv) a dense compact ($\sim$1.5$''$) molecular core surrounding the star and slowly ($\le$12~km~s$^{-1}$) expanding;
v) a compact (about $0.2\times0.4''$) HII region close to the central object, indicating the onset of ionisation in the envelope.

When the central star of a PPN becomes hot enough to ionize its CSE (T$_{\rm eff} \ge $20$\,000$--30$\,000$~K), radio continuum emission from the ionized gas  can be detected in the centimetric range and this has been used to search for hot post-AGB stars where the ionisation has recently started \citep{umana, cerrigone11, cerrigone17}.
CRL~618 was first detected at radio wavelengths by \citet{wynn} and was soon found to display increasing flux density over time, which was interpreted as due to the expansion of the ionized region \citep{kwok81}. Since then, several works have addressed the variability of the radio flux from this object, which displays 
 a steady increase in the optically thick centimetric range and a more erratic behavior  in the millimeter range, where the emission is optically thin \citep{martin88, sanchez04_1, planck, sanchezIRAM}. While the increase in flux density at frequencies where the emission is optically thick is interpreted as the expansion of the emitting surface, the erratic variability at frequencies of optically thin emission has been seen as an indication of on-going activity from a stellar post-AGB wind \citep{sanchezIRAM}.

 CRL~618 is also the only proto-PN studied in millimeter radio recombination lines (RRLs), which has allowed estimating a remarkable mass-loss rate of $\sim$8.4$\times 10^{-6}$~M$_\odot$~yr$^{-1}$ for its post-AGB wind \citep{martin88,sanchezIRAM}. The existence of this still active ionized wind from the central star has also been invoked by \citet{tafoya13}, who analyzed radio data of CRL~618 spanning years 1982--2007 and concluded that the ionisation of the circumstellar material started around 1971. \citet{tafoya13} also confirmed that the nebula is ionisation-bounded in the direction of its minor axis, indicating a much larger density in the equatorial direction than along the polar axis.

The distance to CRL~618 is still somewhat uncertain. \citet{schmidt1981} estimated a distance of 1.8~kpc, assuming a main sequence luminosity of $2\times10^4$~L$_\odot$ for the B0 central star. \citet{goodrich1991} argued instead for a smaller luminosity of $10^4$~L$_\odot$, hence a distance of 0.9~kpc, although based on the radial velocity of the star and the Galactic rotation curve, they also mentioned a possible value of 3.1~kpc. \citet{knapp1993} estimated a bolometric flux of $2\times10^{-7}$~erg~cm$^{-2}$~s$^{-1}$ and quoted a distance of 1.1~kpc, for an assumed luminosity of  $10^4$~L$_\odot$. Finally, \citet{sanchez04_1} measured proper motions in the high-velocity bipolar clumps that support the value of 0.9~kpc indicated by \citet{goodrich1991}.

In the context of estimating the distance to a star, geometrical methods are the most reliable ones, when possible. One of these is the so-called expansion parallax, which allows deriving the distance by measuring the angular expansion of a structure such as a ring or shell, if its linear expansion velocity is known. Since the development of a PN implies the existence and expansion of an ionized shell, this method can be applied to this kind of sources even in their early phases, if their angular expansion and linear velocity can be measured. A version of this method was developed by \citet{masson} based on the comparison of radio interferometric visibility data sets and was later applied to several PNe \citep[and references therein]{guzmanNGC6881}.

\section{Observations and data reduction}
We observed CRL~618 with the NSF Karl G. Jansky Very Large Array (VLA) telescope operated by the National Radio Astronomy Observatory (NRAO), while the array was in configuration A on 2011 June 12, at a frequency of $\sim$43~GHz (7~mm). The data were acquired within project 11A-171 (PI: G. Umana). The flux calibrator was 3C~138 and the complex-gain calibrator was JVAS~J0414+3418. Two spectral windows were set up, spanning a total of 64~MHz in full polarization. 
The duration of the observations was 2 hours. 

The data were reduced in CASA 5.6.1-8 \citep{casa} with the VLA pipeline delivered with the software package, after correcting the  {intents\footnote{Intents are used by the pipeline to identify calibrators (in our case, pointing, bandpass, flux, and complex-gain calibrators) and science targets in an observation.}} of the astrophysical sources in it:  {3C~138 turned out not to have sufficient signal-to-noise ratio per channel to be used as bandpass calibrator, hence it was used only as flux calibrator}, while $J0414+3418$ was used as both bandpass and complex-gain calibrator. After a first run of the pipeline, the data were inspected and some necessary manual flags were identified. The whole calibration was then repeated re-starting from the raw data and including the manual flags. Given the small band width of the data set, no correction was performed for the spectral slope introduced using the complex-gain calibrator to calibrate the bandpass. 

In this paper, we also make use of the data taken with the VLA on 1998-05-02 within project AW485 (PI: J. Wrobel). The data were acquired in continuum mode and span 100~MHz in frequency, centered around 43.3~GHz in full polarization. The array was in A configuration, but only a subarray of 12 antennas was used at 43~GHz, while a simultaneous subarray was used to observe in different frequency bands. The duration of the observation was about 8~hours.

This second data set was calibrated setting explicitly the flux scale of the phase calibrator $\mathrm{JVAS}\;0443+3441$ to the same values reported by \citet{tafoya13}\footnote{Please, notice that project AW485 is reported as AW048 by \citet{tafoya13}.}, since the flux calibrator in the original data is now known to be variable, hence not reliable. {It must be noticed that the critical parameter for the analysis carried out in this work is not the absolute flux calibration in each data set, but the size of the emitting area, which is identified by a signal-to-noise ratio larger than 5, therefore it does not depend on the absolute calibration.} Opacity and gain-curve correction were applied to the data, then a phase-only calibration table was generated, which was applied to generate the final amplitude and phase calibration. {The reduction of the AW485 data was performed with the same CASA version used for the 11A-171 data set.}

Both the 1998 and the 2011 data sets were self-calibrated only in phase, after the initial phase-reference calibration. This led to final flux densities of 0.98$\,\pm\,$0.05~Jy~beam$^{-1}$ and 1.20$\,\pm\,$0.06 Jy~beam$^{-1}$ in 1998 and 2011, respectively. {The rms noise in each of the maps was instead of 0.4~mJy~beam$^{-1}$ in 11A-171 and 0.8~mJy~beam$^{-1}$ in AW485.}

\begin{figure}
    \centering
    \includegraphics[width=0.45\textwidth]{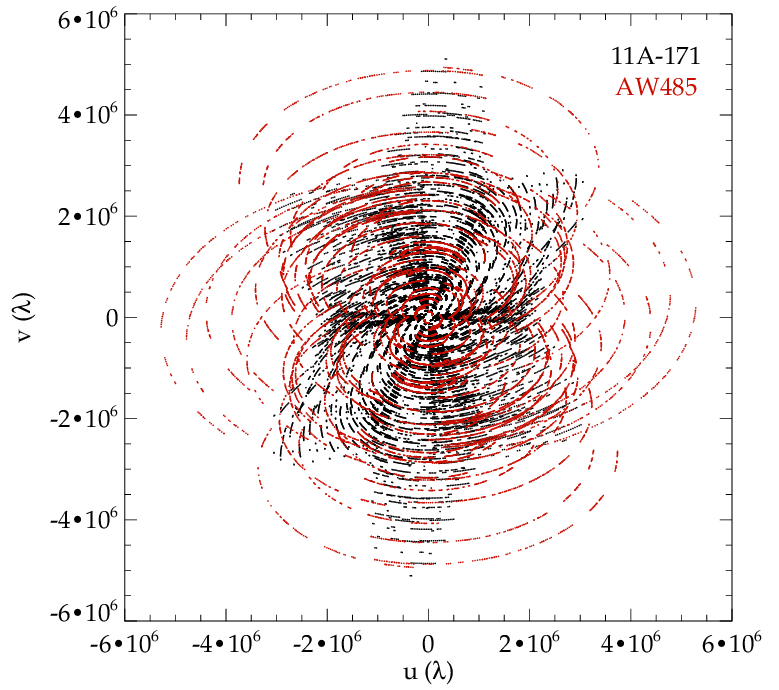}
    \caption{$u\varv$ coverage of data sets 11A-171 and AW485  after calibration and flagging.}
    \label{fig:uvwave}
\end{figure}

\begin{figure}
    \centering
    \includegraphics[width=0.45\textwidth]{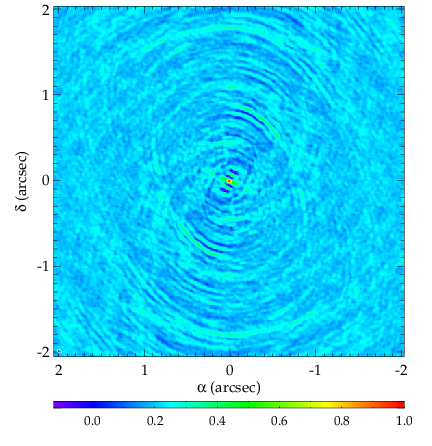}\qquad
    \includegraphics[width=0.45\textwidth]{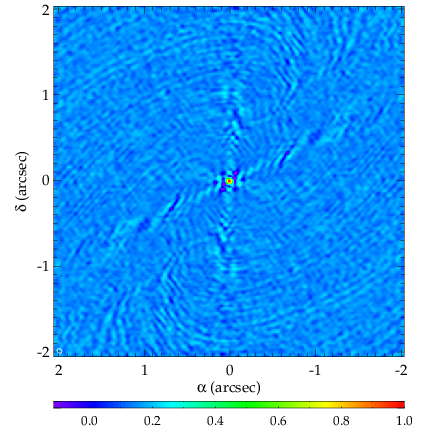}
    \caption{Dirty beams of the AW485 (\textit{top}) and 11A-171 (\textit{bottom}) data sets.}
    \label{fig:dirtybeams}
\end{figure}

In Figure~\ref{fig:uvwave}, we display the final $u\varv$ coverage for CRL~618 in the two data sets at 43~GHz. It is evident that the sampling is less dense in AW485 due to the smaller number of antennas, but at the same time, the longer integration time allows for a more uniform distribution over the plane. AW485 achieves about the same baseline lengths in both ${u}$ and {$\varv$}, which is expected to turn into a rounder synthesized beam. The  11A-171 data set achieves about the same baseline lengths in $\varv$ but  shorter in \textit{u}. 
Although the $u\varv$ sampling is clearly different in the two data sets, these differences do not seem to be so substantial as to bias the final imaging products and their analysis{, especially for values of $|u|$ and $|\varv|$ smaller than 3.5~M$\lambda$, which corresponds to a beam size of $\sim$0.06$''$. The differences can then be well} mitigated through data weighting and beam convolution in the imaging step. {In Figure~\ref{fig:dirtybeams}, we display the dirty beams of the two data sets.}

The 80th percentiles  for the shortest and longest baselines in the two final data sets are respectively 3744~m and 15407~m in 11A-171 and 5377~m and 20408~m in AW485. 
The sampling in AW485 is therefore shifted towards longer baselines in comparison to 11A-171, due to the longer integration and different antenna distribution. To overcome the different angular resolution, we  convolved the final products to a common beam with size matching both data sets. 

\section{Nebular expansion}
We performed our analysis of the expansion of the circumstellar ionized nebula in CRL~618 with the AW485 and 11A-171 data sets at 43 GHz following the method described by \citet{masson}, to obtain a difference image from two radio interferometric data sets. 

As explained above, the data sets were initially calibrated independently, then a round of self-calibration in amplitude and phase was executed on 11A-171, assuming as a model the image from AW485.  {As required by the radio parallax method \citep{masson}}, this was done to align the two images on the same amplitude and phase reference, removing any offsets.

\begin{figure}
    \centering
    \includegraphics[scale=0.45]{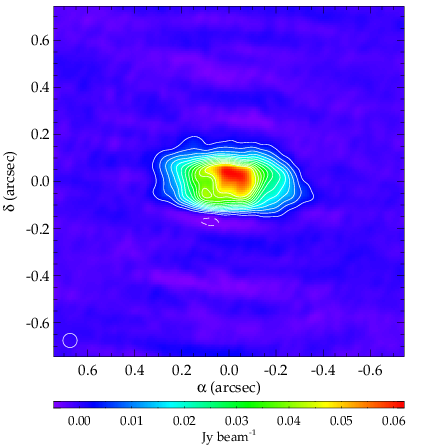} %\qquad \qquad
    \includegraphics[scale=0.45]{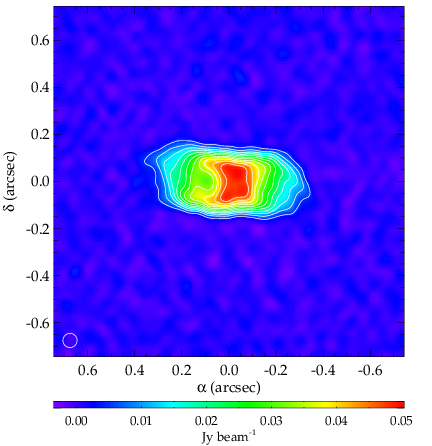} 
    \includegraphics[scale=0.45]{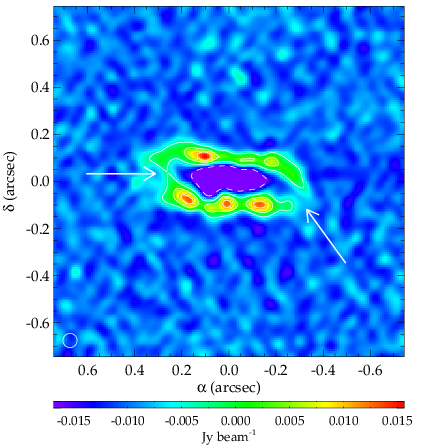}
    \caption{Images of CRL~618 in Q band from projects AW485 (top) and 11A-171 (middle); contours start at 5$\times$rms and increase by steps of 5$\times$rms ($-5\times$rms is also plotted as a dashed line); the convolution beam of 0.06$''$ is plotted in the bottom left; {the rms noise is 0.8 and 0.4 mJy~beam$^{-1}$ for the AW485 and 11A-171 data, respectively.} The bottom image displays the difference of the two data sets, linearly stretched between $-5\times$rms and its maximum{, with two arrows pointing at drops in flux density along the shell}.}
    \label{fig:Q_imaging}    
\end{figure}

 {After this step, the AW485 data  were subtracted in the visibility plane from 11A-171, using the CASA task \textit{uvsub}}. At this point, imaging of the difference data set was performed with \textit{tclean} in CASA 5.6.1-8. In Figure~\ref{fig:Q_imaging}, we display the images of the source at the two epochs and the difference obtained. 
The width of the {convolution} beam was set to 0.06${''}$ for both data sets.

{The difference image (the bottom plane in Figure~\ref{fig:Q_imaging}) displays a clear ellipse of emission.} Along the direction of the major axis {of the ellipse}, two drops in emission at the eastern-most and western-most locations can be seen {(indicated by two arrows in the figure)}.  {This is compatible with the known difference in density between the material along the polar axis, which has been mostly swept away by the fast wind, and the material in the  perpendicular direction, which is still an almost unaltered residual from the AGB phase} \citep{sanchez02}. The inner region has negative emission (i.e., the older image has larger flux density in that area), as expected due to the expansion. 

\subsection{Expansion-parallax distance}
After detecting the expansion in the difference image, we estimated the distance to our target as described by \citet{guzmanNGC6881}.

We started from the image of CRL~618 obtained with the AW485 data set, then created a set of expanded images, each of them with an expansion factor of $1+\epsilon$, where $\epsilon$ ranged from 0.1 to 0.3 in steps of 0.003. {From every expanded image we then subtracted the AW485 one}, resulting into a grid of model difference images. Every model difference image thus obtained was compared to the real difference image  pixel by pixel, calculating a $\chi^2$ value

\begin{equation}
    \chi^2 = \sum_{\rm ij} \frac{(M_{\rm ij}-T_{\rm ij})^2}{\sigma^2}
    \label{eq:chi}
\end{equation}
where $M_{\rm ij}$ is the value of the pixel with coordinates $i$ and $j$ in the model difference and $T_{\rm ij}$ in the real difference, and $\sigma$ is the standard deviation of the $M_{\rm ij}-T_{\rm ij}$ values  {over the whole map}. 

\begin{figure}
    \centering    
    \includegraphics[width=0.45\textwidth]{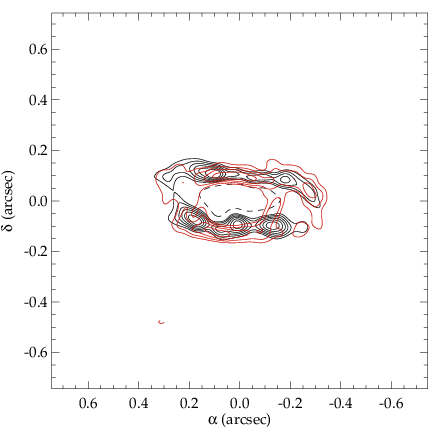}
    \caption{Real (black contours) and best-model (red contours) difference images, with contour levels at rms$\times(-5, 5, 7, 9, 11, 13, 15)$ and rms of about 0.8~mJy~beam$^{-1}$ and 1~mJy~beam$^{-1}$, respectively. Negative levels are displayed as dashed lines and positive ones as solid lines.}
    \label{fig:comparediff}
\end{figure}

{For a direct comparison, we display in Figure~\ref{fig:comparediff} the contours of the real difference image in black and those of the best-model difference image in red.}

\begin{figure}
    \centering
    \includegraphics[width=0.4\textwidth]{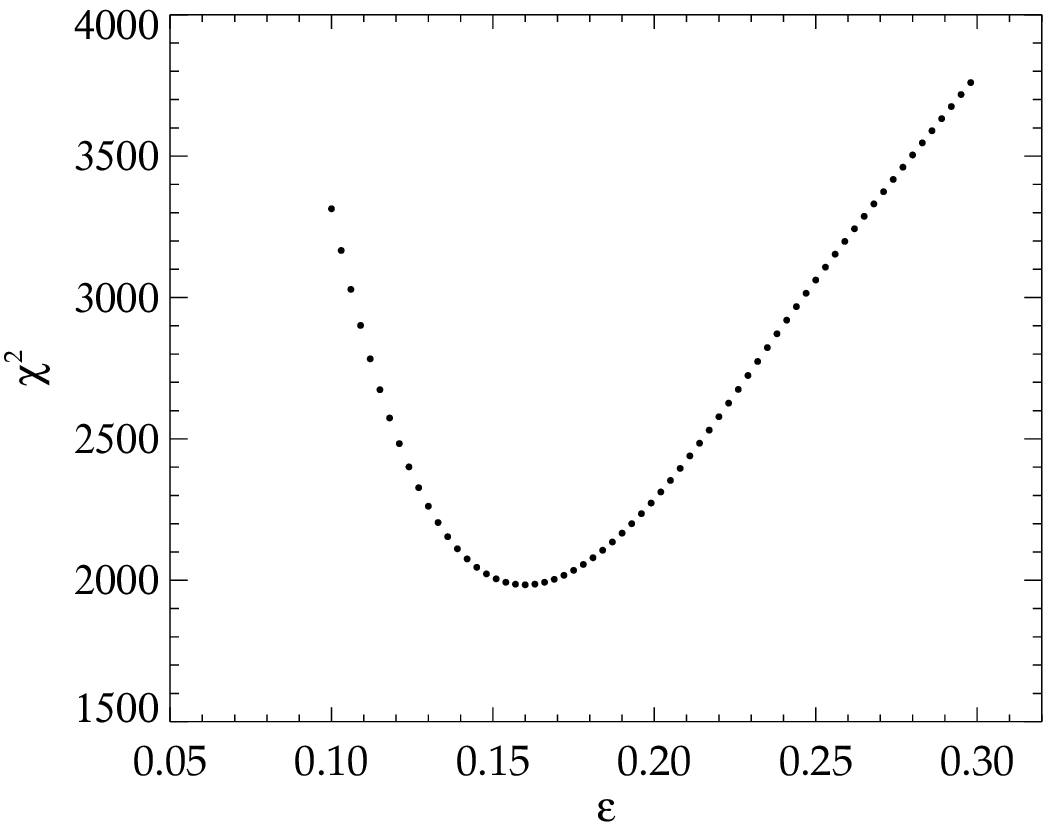}
    \caption{Distribution of $\chi^2$ (as defined in Eq.~\ref{eq:chi}) as a function of $\epsilon$ (the expansion factor minus one).}
    \label{fig:chisq}
\end{figure}

\begin{figure}
    \centering
    \includegraphics[width=0.4\textwidth]{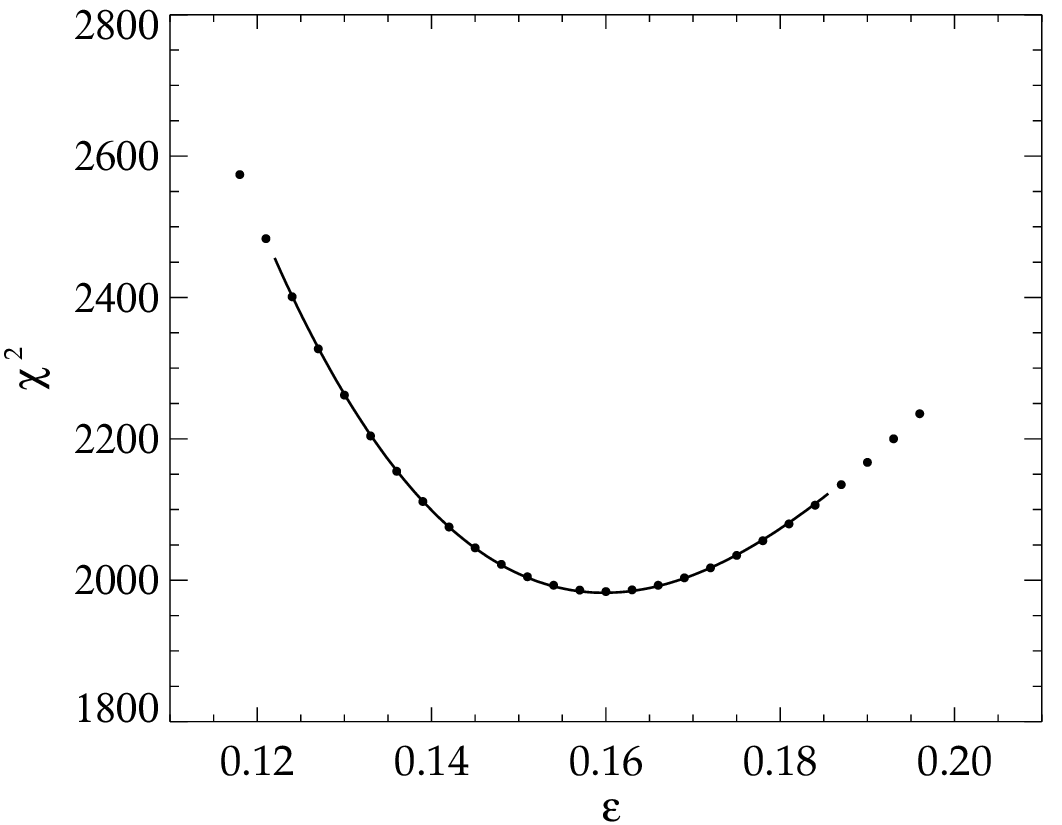}
    \caption{Cubic curve fitted to the $\chi^2$ values in the range $0.2 \le \epsilon \le 0.225$.}
    \label{fig:chimin}
\end{figure}

The values of $\chi^2$ obtained are plotted in Figure~\ref{fig:chisq} as a function of $\epsilon$. The minimum was found by fitting a cubic curve in the range where $0.12 \le \epsilon \le 0.2$ (Figure~\ref{fig:chimin})
and then calculating its analytical minimum. $\chi^2$ was then found to be minimum at $\epsilon = 0.160 \pm 0.003$, where the error is the statistical error on the fit to the curve of $\chi^2$ values.

The distance to the object can be calculated with the following formula \citep{guzmanNGC6881}:
\begin{equation}
    \left[\frac{d}{\rm pc}\right] = 211 \, \left[\frac{\varv_{{\rm exp}}}{\rm km\;s^{-1}}\right] \, \left[\frac{\dot{\theta}}{\rm mas \; yr^{-1}}\right]^{-1}
    \label{eq:distance}
\end{equation}
where $\varv_{\rm exp}$ is the expansion velocity and $\dot{\theta}$ is the angular expansion rate in milliarcsec (mas) per year, which is equivalent to

\begin{equation}
    \frac{\rm \epsilon \, \theta_{mas} }{13.1205 \;\; \rm yr}
    \label{eq:easy_dist}
\end{equation} 

where $\epsilon$ is the expansion factor that we have determined, $\theta$ the angular radius at the time of the AW485 observation in mas, and 13.1205~yr is the time between the AW485 and 11A-171 observations.

To obtain an estimation of the size of the nebula in 1998, we first rotated the AW485 image  by 2.5 degrees in the direction North to East, to align its main axis with the E-W direction, then slices of the nebula were taken in the N-S and E-W directions separately, measuring the length of every slice (amount of pixels above 5$\sigma$ multiplied by pixel length) and summing the flux density values along each of them. Finally, a flux-weighted mean size was calculated as $$\frac{\sum_il_i \, S_i}{\sum_j S_j}$$ where $l$ is the length of every slice and $S$ the total flux from it. This returned a size of $0.34\arcsec\times0.65\arcsec$, with a relative error of about 10\%, considering the error of 5\% on the flux density measurements. We can then use the length of the  semi-axes to derive the angular expansion along the polar direction and orthogonally to it.  
This translates according to Eq.\ref{eq:easy_dist} into $\dot{\theta} = 2.1 \pm 0.2 \; $mas~yr$^{-1}$ along the minor axis and $4.0 \pm 0.4 \; $mas~yr$^{-1}$ along the major axis, between years 1998 and 2011. These compare well with the values of $2.3 \pm 0.6\; $mas~yr$^{-1}$ and $4.7 \pm 1.1  \; $mas~yr$^{-1}$ found by \citet{tafoya13} and averaged over time until 2007.

In the derivation of the distance from Eq.~\ref{eq:distance}, a critical value is the velocity. It has been shown that expansion-parallax distances need to be corrected, if the velocity is obtained from optical spectral lines, due to the different velocities of the material traced by the radio continuum and the lines of ionized elements \citep{schoenberner}. To avoid this complication, we make use of the velocity derived from radio data by \citet{martin88} and \citet{sanchezIRAM}, who estimate that the HII region expands at $\sim$20~km~s$^{-1}$ by modelling its radio recombination lines (namely, H30$\alpha$, H35$\alpha$, and H41$\alpha$), thus tracing the overall motion of the ionized gas, without abundance or line-excitation biases. As the expansion occurs mainly along the major axis, we associate this velocity (with a 10\% error) to the  expansion rate in this direction and thus obtain a distance of $1.1\pm0.2$~kpc. Though slightly larger, this value compares well with that of  $\sim$900~pc derived from the expansion in the lobes, seen in the optical images of CRL~618 \citep{sanchez04_1}.

\section{Conclusions}
We have analyzed data of CRL~618 obtained in 1998 and 2011 at 43~GHz (7~mm) with the VLA. The 2011 data are presented here for the first time. A size increase of the nebula is immediately evident by eye in the images from the different epochs. The expansion was estimated by imaging the data after subtraction in the visibility plane and comparing this to a grid of model difference images obtained expanding the image from 1998. This assumes that the expansion is self-similar, which is not strictly true for CRL~618, because its major and minor axes have been found to expand at different rates. Nevertheless, our analysis indicates  that at first approximation this can be neglected and the method still returns reliable results. The expansion rates that we find are in fact compatible within errors with previous independent estimations: we find $2.1\pm0.2$~mas~yr$^{-1}$ along the minor axis and $4.0\pm0.4$~mas~yr$^{-1}$ along the major axis, leading to a distance to CRL~618 of $1.1\pm0.2$~kpc. While previous estimations of the distance to CRL~618 were derived from an assumed intrinsic luminosity of the source, this is its first direct measurement. The present value matches within errors with  that of 0.9~kpc, which has been widely adopted in the last twenty years, and coincides with that given by \citet{knapp1993} for a luminosity of 10$^4$~L$_\odot$.

\section*{Acknowledgements}

The National Radio Astronomy Observatory is a facility of the National Science Foundation operated under cooperative agreement by Associated Universities, Inc.

%%%%%%%%%%%%%%%%%%%%%%%%%%%%%%%%%%%%%%%%%%%%%%%%%%
\section*{Data Availability}
The data underlying this article were accessed from \url{https://data.nrao.edu/}, under project codes 11A-171 and AW485. The derived data generated in this research will be shared on reasonable request to the corresponding author.



\bibliographystyle{mnras}
\bibliography{crl618} % if your bibtex file is called example.bib

\begin{thebibliography}{}
\makeatletter
\relax
\def\mn@urlcharsother{\let\do\@makeother \do\$\do\&\do\#\do\^\do\_\do\%\do\~}
\def\mn@doi{\begingroup\mn@urlcharsother \@ifnextchar [ {\mn@doi@}
  {\mn@doi@[]}}
\def\mn@doi@[#1]#2{\def\@tempa{#1}\ifx\@tempa\@empty \href
  {http://dx.doi.org/#2} {doi:#2}\else \href {http://dx.doi.org/#2} {#1}\fi
  \endgroup}
\def\mn@eprint#1#2{\mn@eprint@#1:#2::\@nil}
\def\mn@eprint@arXiv#1{\href {http://arxiv.org/abs/#1} {{\tt arXiv:#1}}}
\def\mn@eprint@dblp#1{\href {http://dblp.uni-trier.de/rec/bibtex/#1.xml}
  {dblp:#1}}
\def\mn@eprint@#1:#2:#3:#4\@nil{\def\@tempa {#1}\def\@tempb {#2}\def\@tempc
  {#3}\ifx \@tempc \@empty \let \@tempc \@tempb \let \@tempb \@tempa \fi \ifx
  \@tempb \@empty \def\@tempb {arXiv}\fi \@ifundefined
  {mn@eprint@\@tempb}{\@tempb:\@tempc}{\expandafter \expandafter \csname
  mn@eprint@\@tempb\endcsname \expandafter{\@tempc}}}

\bibitem[\protect\citeauthoryear{{Balick}, {Riera}, {Raga}, {Kwitter}  \&
  {Vel{\'a}zquez}}{{Balick} et~al.}{2014}]{balick14}
{Balick} B.,  {Riera} A.,  {Raga} A.,  {Kwitter} K.~B.,   {Vel{\'a}zquez}
  P.~F.,  2014, \mn@doi [\apj] {10.1088/0004-637X/795/1/83}, \href
  {https://ui.adsabs.harvard.edu/abs/2014ApJ...795...83B} {795, 83}

\bibitem[\protect\citeauthoryear{{Cerrigone}, {Trigilio}, {Umana}, {Buemi}  \&
  {Leto}}{{Cerrigone} et~al.}{2011}]{cerrigone11}
{Cerrigone} L.,  {Trigilio} C.,  {Umana} G.,  {Buemi} C.~S.,   {Leto} P.,
  2011, \mn@doi [\mnras] {10.1111/j.1365-2966.2010.17968.x}, \href
  {https://ui.adsabs.harvard.edu/abs/2011MNRAS.412.1137C} {412, 1137}

\bibitem[\protect\citeauthoryear{{Cerrigone}, {Umana}, {Trigilio}, {Leto},
  {Buemi}  \& {Ingallinera}}{{Cerrigone} et~al.}{2017}]{cerrigone17}
{Cerrigone} L.,  {Umana} G.,  {Trigilio} C.,  {Leto} P.,  {Buemi} C.~S.,
  {Ingallinera} A.,  2017, \mn@doi [\mnras] {10.1093/mnras/stx690}, \href
  {https://ui.adsabs.harvard.edu/abs/2017MNRAS.468.3450C} {468, 3450}

\bibitem[\protect\citeauthoryear{{De Marco} \& {Izzard}}{{De Marco} \&
  {Izzard}}{2017}]{demarco}
{De Marco} O.,  {Izzard} R.~G.,  2017, \mn@doi [\pasa] {10.1017/pasa.2016.52},
  \href {https://ui.adsabs.harvard.edu/abs/2017PASA...34....1D} {34, e001}

\bibitem[\protect\citeauthoryear{{Goodrich}}{{Goodrich}}{1991}]{goodrich1991}
{Goodrich} R.~W.,  1991, \mn@doi [\apj] {10.1086/170313}, \href
  {https://ui.adsabs.harvard.edu/abs/1991ApJ...376..654G} {376, 654}

\bibitem[\protect\citeauthoryear{{Guzm{\'a}n-Ram{\'\i}rez}, {G{\'o}mez},
  {Loinard}  \& {Tafoya}}{{Guzm{\'a}n-Ram{\'\i}rez}
  et~al.}{2011}]{guzmanNGC6881}
{Guzm{\'a}n-Ram{\'\i}rez} L.,  {G{\'o}mez} Y.,  {Loinard} L.,   {Tafoya} D.,
  2011, \mn@doi [\mnras] {10.1111/j.1365-2966.2011.18609.x}, \href
  {https://ui.adsabs.harvard.edu/abs/2011MNRAS.414.3129G} {414, 3129}

\bibitem[\protect\citeauthoryear{{Knapp}, {Sandell}  \& {Robson}}{{Knapp}
  et~al.}{1993}]{knapp1993}
{Knapp} G.~R.,  {Sandell} G.,   {Robson} E.~I.,  1993, \mn@doi [\apjs]
  {10.1086/191820}, \href
  {https://ui.adsabs.harvard.edu/abs/1993ApJS...88..173K} {88, 173}

\bibitem[\protect\citeauthoryear{{Kwok} \& {Feldman}}{{Kwok} \&
  {Feldman}}{1981}]{kwok81}
{Kwok} S.,  {Feldman} P.~A.,  1981, \mn@doi [\apjl] {10.1086/183591}, \href
  {https://ui.adsabs.harvard.edu/abs/1981ApJ...247L..67K} {247, L67}

\bibitem[\protect\citeauthoryear{{Kwok}, {Purton}  \& {Fitzgerald}}{{Kwok}
  et~al.}{1978}]{kwok78}
{Kwok} S.,  {Purton} C.~R.,   {Fitzgerald} P.~M.,  1978, \mn@doi [\apjl]
  {10.1086/182621}, \href
  {https://ui.adsabs.harvard.edu/abs/1978ApJ...219L.125K} {219, L125}

\bibitem[\protect\citeauthoryear{{Martin-Pintado}, {Bujarrabal}, {Bachiller},
  {Gomez-Gonzalez}  \& {Planesas}}{{Martin-Pintado} et~al.}{1988}]{martin88}
{Martin-Pintado} J.,  {Bujarrabal} V.,  {Bachiller} R.,  {Gomez-Gonzalez} J.,
  {Planesas} P.,  1988, \aap, \href
  {https://ui.adsabs.harvard.edu/abs/1988A&A...197L..15M} {197, L15}

\bibitem[\protect\citeauthoryear{{Masson}}{{Masson}}{1986}]{masson}
{Masson} C.~R.,  1986, \mn@doi [\apjl] {10.1086/184630}, \href
  {https://ui.adsabs.harvard.edu/abs/1986ApJ...302L..27M} {302, L27}

\bibitem[\protect\citeauthoryear{{McMullin}, {Waters}, {Schiebel}, {Young}  \&
  {Golap}}{{McMullin} et~al.}{2007}]{casa}
{McMullin} J.~P.,  {Waters} B.,  {Schiebel} D.,  {Young} W.,   {Golap} K.,
  2007, in {Shaw} R.~A.,  {Hill} F.,   {Bell} D.~J.,  eds,  Astronomical
  Society of the Pacific Conference Series Vol. 376, Astronomical Data Analysis
  Software and Systems XVI. p.~127

\bibitem[\protect\citeauthoryear{{Pardo}, {Cernicharo}, {Goicoechea}  \&
  {Phillips}}{{Pardo} et~al.}{2004}]{pardo}
{Pardo} J.~R.,  {Cernicharo} J.,  {Goicoechea} J.~R.,   {Phillips} T.~G.,
  2004, \mn@doi [\apj] {10.1086/424379}, \href
  {https://ui.adsabs.harvard.edu/abs/2004ApJ...615..495P} {615, 495}

\bibitem[\protect\citeauthoryear{{Planck Collaboration} et~al.,}{{Planck
  Collaboration} et~al.}{2015}]{planck}
{Planck Collaboration} et~al., 2015, \mn@doi [\aap]
  {10.1051/0004-6361/201423836}, \href
  {https://ui.adsabs.harvard.edu/abs/2015A&A...573A...6P} {573, A6}

\bibitem[\protect\citeauthoryear{{S{\'a}nchez Contreras} \&
  {Sahai}}{{S{\'a}nchez Contreras} \& {Sahai}}{2004}]{sanchez04_1}
{S{\'a}nchez Contreras} C.,  {Sahai} R.,  2004, \mn@doi [\apj]
  {10.1086/381179}, \href
  {https://ui.adsabs.harvard.edu/abs/2004ApJ...602..960S} {602, 960}

\bibitem[\protect\citeauthoryear{{S{\'a}nchez Contreras}, {Sahai}  \& {Gil de
  Paz}}{{S{\'a}nchez Contreras} et~al.}{2002}]{sanchez02}
{S{\'a}nchez Contreras} C.,  {Sahai} R.,   {Gil de Paz} A.,  2002, \mn@doi
  [\apj] {10.1086/342316}, \href
  {https://ui.adsabs.harvard.edu/abs/2002ApJ...578..269S} {578, 269}

\bibitem[\protect\citeauthoryear{{S{\'a}nchez Contreras}, {Bujarrabal},
  {Castro-Carrizo}, {Alcolea}  \& {Sargent}}{{S{\'a}nchez Contreras}
  et~al.}{2004}]{sanchez04_2}
{S{\'a}nchez Contreras} C.,  {Bujarrabal} V.,  {Castro-Carrizo} A.,  {Alcolea}
  J.,   {Sargent} A.,  2004, \mn@doi [\apj] {10.1086/425409}, \href
  {https://ui.adsabs.harvard.edu/abs/2004ApJ...617.1142S} {617, 1142}

\bibitem[\protect\citeauthoryear{{S{\'a}nchez Contreras}, {B{\'a}ez-Rubio},
  {Alcolea}, {Bujarrabal}  \& {Mart{\'\i}n-Pintado}}{{S{\'a}nchez Contreras}
  et~al.}{2017}]{sanchezIRAM}
{S{\'a}nchez Contreras} C.,  {B{\'a}ez-Rubio} A.,  {Alcolea} J.,  {Bujarrabal}
  V.,   {Mart{\'\i}n-Pintado} J.,  2017, \mn@doi [\aap]
  {10.1051/0004-6361/201730385}, \href
  {https://ui.adsabs.harvard.edu/abs/2017A&A...603A..67S} {603, A67}

\bibitem[\protect\citeauthoryear{{Schmidt} \& {Cohen}}{{Schmidt} \&
  {Cohen}}{1981}]{schmidt1981}
{Schmidt} G.~D.,  {Cohen} M.,  1981, \mn@doi [\apj] {10.1086/158943}, \href
  {https://ui.adsabs.harvard.edu/abs/1981ApJ...246..444S} {246, 444}

\bibitem[\protect\citeauthoryear{{Sch{\"o}nberner}, {Balick}  \&
  {Jacob}}{{Sch{\"o}nberner} et~al.}{2018}]{schoenberner}
{Sch{\"o}nberner} D.,  {Balick} B.,   {Jacob} R.,  2018, \mn@doi [\aap]
  {10.1051/0004-6361/201731788}, \href
  {https://ui.adsabs.harvard.edu/abs/2018A&A...609A.126S} {609, A126}

\bibitem[\protect\citeauthoryear{{Soker}}{{Soker}}{1998}]{soker1998}
{Soker} N.,  1998, \mn@doi [\apj] {10.1086/305407}, \href
  {https://ui.adsabs.harvard.edu/abs/1998ApJ...496..833S} {496, 833}

\bibitem[\protect\citeauthoryear{{Soria-Ruiz}, {Bujarrabal}  \&
  {Alcolea}}{{Soria-Ruiz} et~al.}{2013}]{soria}
{Soria-Ruiz} R.,  {Bujarrabal} V.,   {Alcolea} J.,  2013, \mn@doi [\aap]
  {10.1051/0004-6361/201322133}, \href
  {https://ui.adsabs.harvard.edu/abs/2013A&A...559A..45S} {559, A45}

\bibitem[\protect\citeauthoryear{{Tafoya}, {Loinard}, {Fonfr{\'\i}a},
  {Vlemmings}, {Mart{\'\i}-Vidal}  \& {Pech}}{{Tafoya} et~al.}{2013}]{tafoya13}
{Tafoya} D.,  {Loinard} L.,  {Fonfr{\'\i}a} J.~P.,  {Vlemmings} W.~H.~T.,
  {Mart{\'\i}-Vidal} I.,   {Pech} G.,  2013, \mn@doi [\aap]
  {10.1051/0004-6361/201321704}, \href
  {https://ui.adsabs.harvard.edu/abs/2013A&A...556A..35T} {556, A35}

\bibitem[\protect\citeauthoryear{{Umana}, {Cerrigone}, {Trigilio}  \&
  {Zappal{\`a}}}{{Umana} et~al.}{2004}]{umana}
{Umana} G.,  {Cerrigone} L.,  {Trigilio} C.,   {Zappal{\`a}} R.~A.,  2004,
  \mn@doi [\aap] {10.1051/0004-6361:200400059}, \href
  {https://ui.adsabs.harvard.edu/abs/2004A&A...428..121U} {428, 121}

\bibitem[\protect\citeauthoryear{{Westbrook}, {Becklin}, {Merrill},
  {Neugebauer}, {Schmidt}, {Willner}  \& {Wynn-Williams}}{{Westbrook}
  et~al.}{1975}]{westbrook}
{Westbrook} W.~E.,  {Becklin} E.~E.,  {Merrill} K.~M.,  {Neugebauer} G.,
  {Schmidt} M.,  {Willner} S.~P.,   {Wynn-Williams} C.~G.,  1975, \mn@doi
  [\apj] {10.1086/153989}, \href
  {https://ui.adsabs.harvard.edu/abs/1975ApJ...202..407W} {202, 407}

\bibitem[\protect\citeauthoryear{{Wynn-Williams}}{{Wynn-Williams}}{1977}]{wynn}
{Wynn-Williams} C.~G.,  1977, \mn@doi [\mnras] {10.1093/mnras/181.1.61P}, \href
  {https://ui.adsabs.harvard.edu/abs/1977MNRAS.181P..61W} {181, 61P}

\makeatother
\end{thebibliography}

%\begin{thebibliography}{99}
%\end{thebibliography}

%%%%%%%%%%%%%%%%%%%%%%%%%%%%%%%%%%%%%%%%%%%%%%%%%%

%%%%%%%%%%%%%%%%% APPENDICES %%%%%%%%%%%%%%%%%%%%%

%\appendix



%%%%%%%%%%%%%%%%%%%%%%%%%%%%%%%%%%%%%%%%%%%%%%%%%%

% Don't change these lines
\bsp	% typesetting comment
\label{lastpage}
\end{document}